\documentclass[usenatbib]{mn2e}
\usepackage[T1]{fontenc}
\usepackage{times}
\usepackage{graphicx}
\usepackage{amsfonts}
\usepackage[totalwidth=520pt, totalheight=680pt, left=15mm,
  letterpaper]{geometry}

\newcommand{\de}{\mathrm{d}} 

\newcommand{\noi}{\noindent}
\newcommand{\hmpc}{\ensuremath{h^{-1}\,\hbox{Mpc}}}
\newcommand{\etal}{{et~al.}}

\newcommand{\bfv}{{\bf v}}

\newcommand{\mrm}{\mathrm{m}}

\newcommand{\be}{\begin{equation}}
\newcommand{\ee}{\end{equation}}
\newcommand{\kms}{\,\mathrm{km}\,\mathrm{s}^{-1}}
\newcommand{\kmsMpc}{\,\mathrm{km}\,\mathrm{s}^{-1}\,\mathrm{Mpc}^{-1}}
\newcommand{\Mpc}{\,\mathrm{Mpc}}

\title[Influence of the Local Void]{Influence of the Local Void\\on   measurements of the clustering dipole}

\author[Bilicki \& Chodorowski]
{Maciej Bilicki\thanks{E-mail: bilicki@camk.edu.pl}
and Micha{\l} J.\ Chodorowski\thanks{E-mail: michal@camk.edu.pl}
\\ N. Copernicus Astronomical Center, Bartycka  18, 00--716 Warsaw, Poland}

\date{\today}

\begin{document}

\maketitle

\begin{abstract}
In measurements of the clustering dipole from all-sky surveys, an important problem is the lack of information about galaxy distribution in the so-called Zone of Avoidance (ZoA). The existence of the Local Void (LV) has a systematic effect on these measurements. If the ZoA is randomly filled with mock galaxies, then the calculated acceleration of the Local Group of galaxies (LG) has a spurious component, resulting from the lack of real galaxies in the intersection of the LV with the ZoA. This component affects both the misalignment angle between the clustering dipole and the CMB dipole, and the inferred value of mean matter density $\Omega_\mrm$. We calculate the amplitude of the spurious acceleration acting on the LG due to the LV. Its value depends on the geometry and size of the LV, as well as on its density contrast. However, under the simplest assumption of the LV being spherical and completely empty, within the linear theory this amplitude amounts only to about $45\kms$ in units of velocity. The resulting change in the misalignment angle is smaller than $1\degr$, and the fractional change in the deduced value of $\Omega_\mrm$ is about 5 per cent. Accounting for observationally indicated elongation of the LV and maintaining the maximising assumption of a complete lack of galaxies inside increases these numbers only moderately. Specifically, the amplitude of the spurious acceleration rises to about $60\kms$, the misalignment angle remains still smaller than $1\degr$, and the fractional change in the deduced value of $\Omega_\mrm$ is enhanced to about 7 per cent. Thus, despite the overall importance of the Local Void for the motion of the Local Group, the influence of the intersection of the LV with the ZoA on \textit{measurements} of the clustering dipole is found to be only a minor systematic effect.
\end{abstract}


\section{Introduction}

One of the currently applied methods to constrain the cosmological parameter of the non-relativistic matter density, $\Omega_\mrm$, is the comparison of the velocity of the Local Group (LG) with its gravitational acceleration, with the use of the linear-theory equation \citep{Pe80}:
\be\label{eq:dipole}
\bmath{v}_\mathrm{LG}=\frac{H_0\,f(\Omega_\mrm )}{4\pi} \int{ \frac{\delta(\bmath{r}')(\bmath{r}'-\bmath{r}) }{|\bmath{r}'-\bmath{r}  |^3}\,\de^3\bmath{r}'}\;,   
\ee 
where $f(\Omega_\mrm )\simeq\Omega_\mrm^{0.55 }$ \citep{Linder}, $\delta\equiv\rho\slash\rho_\mathrm{b}-1$ is the matter density contrast and the integral is performed in principle over the whole Universe.

The left-hand side of Eq. (\ref{eq:dipole}) is known from the dipole anisotropy of the cosmic microwave background temperature distribution, reduced to the centre of mass of the LG: \mbox{$v_\mathrm{CMB}=622\pm35\, \kms$} in the direction \mbox{$(l,b)=(272\degr\pm3\degr,28\degr\pm5\degr)$} (\citealt{Hinsh}; \citealt{CvdB}). The right-hand side, sometimes called the \textit{clustering dipole}, is calculated from galaxy distribution in all-sky surveys (for details see e.g. \citealt{SW}). However, any `all-sky' survey, regardless of the wavelength of observation, will always miss some information in the Zone of Avoidance (i.e. in the Galactic plane and its vicinity) due to obscuration of extragalactic objects by dust, gas and stars of the Milky Way. In order to compute the integral in Eq. (\ref{eq:dipole}), this fact is overcome by artificially filling the ZoA, in a more or less sophisticated way (see e.g. \citealt{La87}; \citealt{Pl89}; \citealt{LBLB}; \citealt{Mall}; \citealt{PH05}; \citealt{Erdogdu}; \citealt{BP06}). Still, irrespective of the method one chooses, such masking of the Galactic Plane will never completely account for any possibly existing, although unknown, large-scale structures behind the Milky Way. On the other hand, it is often claimed that the direction and amplitude of the calculated clustering dipole do not change significantly for different methods of filling the ZoA (e.g. \citealt{La87}; \citealt{Erdogdu}). However, this does not necessarily mean that possible structures obscured by the Galaxy would have no influence on the calculation of the acceleration of the Local Group.

Recently, \citet{LN08} have studied this issue, using the 2MASS Redshift Survey (2MRS, \citealt{Huch}). They obtained an excess peculiar velocity of the LG towards the ZoA and proposed a hidden galaxy or a galaxy cluster as an explanation of the gravitational pull. This suggestion is however in conflict with the current observational knowledge, as already in 2004 there was confidence that all significant nearby large-scale structures behind the ZoA were known and obscuration of a big galaxy was excluded \citep{FaLa}. Moreover, \citet{LN08} claim that a perfect method of filling the ZoA would give `\textit{no discrepancy between the 2MRS   and the direction of the CMB dipole}' (i.e. between the direction of the acceleration and of the velocity of the LG). This statement is clearly incorrect, since there are other sources of the \textit{misalignment   angle} between these vectors, such as the scatter in the mass-to-light ratio (cf. \citealt{Cr07}) and stochasticity in the non-linear relation between the velocity and acceleration of the LG (\citealt{B99}; \citealt{CCK01}; \citealt{CCBC08}). As for the excess peculiar velocity of the LG, \citet{Tully.etal} propose a different explanation, basing on observational data of galactic distances and velocities: motion \textit{away} from the \textit{Local Void} (LV).\footnote{Note that neither the paper by \citet{Tully.etal}, nor any other papers covering the issue of the Local Void, were  mentioned by \citet{LN08}.}

In the present paper we analyse the possible influence of the LV on \textit{measurements} of the clustering dipole. We would like to emphasize that we do not examine the importance of the Local Void for the motion of the Local Group. As was shown e.g. by \citet{Tully.etal}, the push from the LV is a substantial component of the peculiar velocity of the LG. In our study we are only interested in the effect of masking the intersection of the LV and the ZoA for the purpose of calculation of the clustering dipole within linear theory. Our aim is to investigate possible systematics related to the fact that a part of the LV is hidden behind the ZoA.

The paper is organised as follows. In Section \ref{Sec:LV} we shortly describe the Local Void. Section \ref{Sec:Accel} presents our calculations of the `spurious' acceleration induced by random filling of the LV behind the ZoA. Next, Section \ref{Sec:Shift} covers the issue of the directional shift of the clustering dipole due to the discussed effect. Possible corrections of the density parameter measured from density--velocity comparisons are addressed in Section \ref{Sec:Omega}. Finally, in Section \ref{Sec:Summ} we summarize and conclude.

\section{Local Void}
\label{Sec:LV}
The Local Void is a structure first identified in the Nearby Galaxies Atlas \citep{TF87}. It is poorly defined because much of it lies behind the plane of the Milky Way: its centre is located at Galactic coordinates $l=30\degr$, $b=0\degr$ (Tully 2007, private communication). It is nevertheless confirmed (by surveys in different wavelengths) that there is an underabundance, though not a total lack, of galaxies in a very large part of the sky at low redshifts. This empty region begins at the edge of the Local Group, with the so-called \textit{Local Sheet} as a bounding surface (\citealt{Tully08a}; \citealt{Tully08b}; \citealt{Tully.etal}). Both the shape of the LV and its dimensions are not exactly known; however, it seems to have two components: a smaller void with a long dimension of $\sim 35 \, \Mpc$ enclosed within a larger void with a long dimension of the order $60\div70 \, \Mpc$ (\citealt{Tully07}; \citealt{Tully.etal}).

In the framework of cosmological gravitational instability, voids are expected to form out of initial underdensities, i.e. regions less dense than average. Their expansion is faster than the Hubble flow, which results in voids `swelling'. \citet{Tully.etal} calculated an `effective Hubble rate' of the Local Void, under the simplest assumption of it being empty and spherical, and used this calculation, together with the peculiar velocity of the LG away from the LV, to estimate the radius of the latter as at least 16~Mpc. However, these dynamical estimates cannot exclude an effective diameter of the LV even greater than 45~Mpc (\citealt{Tully08a}; \citealt{Tully08b}; \citealt{Tully.etal}).

Such a prominent structure, partly hidden behind the Zone of Avoidance, should have an influence on the calculation of the integral in Eq. (\ref{eq:dipole}). In the following Section, using a simple model, we will analyse the amount of the systematic error it may cause for the estimation of the acceleration of the Local Group.

\section{Acceleration due to (the lack of) the Local Void}
\label{Sec:Accel}

Let us define the acceleration vector $\tilde{\bmath{g}}_\mathrm{LG}$ in some convenient units as 
\be
\tilde{\bmath{g}}_\mathrm{LG}=\int \frac{\delta(\bmath{r}')(\bmath{r}'-\bmath{r}) }{|\bmath{r}'-\bmath{r}  |^3}\,\de^3\bmath{r}'   
\ee 
[cf. Eq. (\ref{eq:dipole})]. We can split the integral in two parts: one over the Local Void and the other covering `the rest of the Universe':
\be
\tilde{\bmath{g}}_\mathrm{LG}=\int\limits_\mathrm{LV} (\ldots) +\int\limits_{\mathbb{R}^3-\mathrm{LV}} (\ldots)\;.   
\ee 
For simplicity, we write only `LV' in this and the following formulae, although we mean in fact `LV$\cap$ZoA', as will be explained later in detail. Now, if we make a simplifying assumption that the Local Void is completely empty ($\rho_\mathrm{LV}=0$), we have $\delta_\mathrm{LV}=-1$. On the other hand, if we filled the part of the Local Void hidden behind the ZoA by randomly chosen `galaxies', i.e. if we assumed average background density in this part ($\rho_\mathrm{LV}=\rho_\mathrm{b} $), we would get $\delta_\mathrm{LV}=0$. We would thus measure some `spurious' acceleration, which is the difference between the calculation made with the ZoA filled randomly and the true peculiar acceleration of the Local Group:
\[
\tilde{\bmath{g}}_\mathrm{spur}=\tilde{\bmath{g}}_\mathrm{cal}-\tilde{\bmath{g}}_\mathrm{T}=\int\limits_\mathrm{LV}{0}\, +\,\int\limits_{\mathbb{R}^3-\mathrm{LV}}{(\ldots)}\,+
\]
\[
\qquad-\,\int\limits_\mathrm{LV}{\frac{(-1)\times(\bmath{r}'-\bmath{r}) }{|\bmath{r}'-\bmath{r}  |^3}\,\de^3\bmath{r}'}\,-\,\int\limits_{\mathbb{R}^3-\mathrm{LV}} (\ldots)=
\]
\be\label{eq:g_spurious}
\qquad=\int\limits_\mathrm{LV} \frac{\bmath{r}'-\bmath{r}}{|\bmath{r}'-\bmath{r}  |^3}\,\de^3\bmath{r}'\;.
\ee
Let us now calculate explicitly the above integral, returning to physical units.

\begin{figure}
\begin{center}
\includegraphics[width=0.45\textwidth]{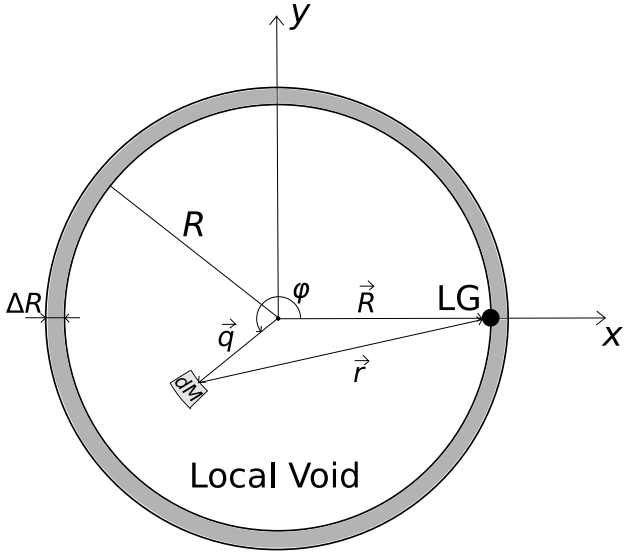}
\caption{\label{Fig:LVxy} A schematic view of the Local Void along the $z$ axis, with the Galactic plane face-on. The relative sizes of the Local Group and the Local Void are not to scale.}
\end{center}
\end{figure}

\begin{figure}
\begin{center}
\includegraphics[width=0.45\textwidth]{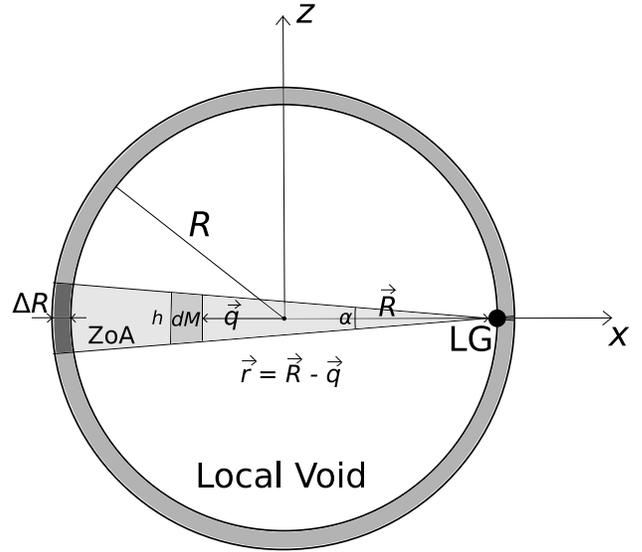} 
\caption{\label{Fig:LVxz} A schematic view of the Local Void along the $y$ axis, with the Galactic plane edge-on. The relative sizes of the Local Group and the Local Void are not to scale.}
\end{center}
\end{figure}

For that purpose, in addition to the assumption of the Local Void being completely empty, we also model it as spherical, with some radius $R$, and assume that the Local Group lies exactly at the edge of the LV (see Figs. \ref{Fig:LVxy} and \ref{Fig:LVxz})\footnote{We neglect the size of the Local Group, estimated usually to about $2\,\hmpc$.}. We place the centre of the Local Void in the plane of the Galactic equator, in accordance with observations, which additionally maximizes the studied effect. The $x$ and $y$ axes of the coordinate system lie in the Galactic plane, with the origin in the centre of the LV (Fig. \ref{Fig:LVxy}). As the latter is situated at $l_\mathrm{LV}=30\degr$, \mbox{$b_\mathrm{LV}=0\degr$}, this means that our coordinate system is shifted with respect to the Galactic one and rotated by $210\degr$ in the Milky Way plane (our $z$ axis is parallel to $z_\mathrm{Gal}$). The part missed in all-sky surveys due to the Zone of Avoidance is a spherical section bounded by two planes (Fig. \ref{Fig:LVxz}); the angle between them, $\alpha$, depends on the survey, but for our purposes $\alpha\simeq 20\degr$ (we consider near-infrared wavelengths, as is explained hereafter). Owing to the smallness of this angle ($\alpha\simeq0.35\,\mathrm{rad}$), we treat the masked region as a thin wedge.

We additionally include a \textit{shell of compensation} at the edge of the LV, of width $\Delta R$. According to the standard picture of void formation, where such structures are grown from initial underdensities and gradually expand, the matter expelled from inside of the void creates a layer at the edge, of density higher than the average background one. This is supported by both simulations (see e.g. \citealt{vdWSch}) and observations (cf. \citealt{Fairall} and references therein): voids are usually bounded by `filaments' and `walls' of high density contrast. In our cosmic neighbourhood, a part of such a shell is probably the Local Sheet and the Local Group is located at its inner edge (Tully 2010, private communication).

A mass element $\de M$ contributing to the spurious acceleration is
\be
\de M=\rho_\mathrm{b}\,\de V = \rho_\mathrm{b}\,h\,\de S = \rho_\mathrm{b}\,\alpha\, r\,\de S\;,
\ee
where $h\simeq\alpha\,r$ is the height of the wedge at a distance $r$ from the LG and $\de S$ is a surface element. The differential acceleration `measured' due to random filling of the ZoA will be
\[
\bmath{\de g}=-G\,\frac{\de M(r)\, \bmath{r}
}{r^3}=-G\,\frac{\rho_\mathrm{b}\,\alpha\,r\,\de S\,\bmath{r} }{r^3}=
\]
\be
\qquad=-\alpha\, G\, \rho_\mathrm{b}\,\frac{q\,\de q \, \de
  \varphi\,(\bmath{R}- \bmath{q}) }{|\bmath{R}- \bmath{q} |^2}\;,
\ee
with $\de S = q\, \de q\, \de \varphi$, $\bmath{q}=(q\cos\varphi,q\sin\varphi) $, $\bmath{R}=(R,0) $ and \mbox{$0\leq\varphi\leq2\pi$} is the azimuthal angle. It is easy to check that from symmetry $g_y=0$. The $x$-component of the spurious acceleration is given by
\be
g_x=-\alpha\,G\,\rho_\mathrm{b}\, R\, \mathcal{I}\;,
\ee
where
\be
\mathcal{I}=\int\limits^{2\pi}_0 \de \varphi\int\limits^1_0 \de \xi \;
\frac{\xi-\xi^2\cos\varphi}{\xi^2+1-2\xi\cos\varphi}=\pi\;.
\ee
Thus the value of the peculiar velocity caused by the spurious acceleration is (in linear theory)
\be\label{eq:v_spurious}
v_\mathrm{spur}
=\frac{H_0\,f(\Omega_\mrm) }{4\pi\,G\,\rho_\mathrm{b}
}\,g=\frac{1}{4} \alpha\,H_0\,f(\Omega_\mrm)\,R\;.
\ee
It is interesting to compare this value with the one induced by a sphere of radius $R$ and density $\rho_\mathrm{b}$. The peculiar acceleration at the surface of the sphere is then $g_\bullet=\frac{4}{3}\pi G \rho_\mathrm{b} R$, which gives the linear peculiar velocity of
\be
v_\bullet=\frac{1}{3}H_0\,f(\Omega_\mrm)\,R\;.
\ee
We thus have
\be
\frac{v_\mathrm{spur}}{v_\bullet}=\frac{3}{4}\, \alpha\;.
\ee

The above calculation applied to an isolated void without any shell of compensation. In order to verify the effect of such a shell, let us first assume that the average matter density inside the layer is constant, $\rho_\mathrm{sh}$, and its thickness is $\Delta R$. The Local Group, of negligible size, is still located at the edge of the Local Void, as in Figures. Putting now the mass element inside the intersection of the shell with the ZoA, we can see that the differential spurious acceleration will read
\be
\bmath{\de g}_\mathrm{sh}=\alpha\, G\, (\rho_\mathrm{sh}-\rho_\mathrm{b})\,\frac{q\,\de q \, \de
  \varphi\,(\bmath{R}- \bmath{q}) }{|\bmath{R}- \bmath{q} |^2}\;,
\ee
with $\bmath{q}$ and $\bmath{R}=(R,0)$ defined as before, but now \mbox{$R\leq q \leq\,R+\Delta R$}. The $y$-component of the acceleration vanishes from symmetry, and along the $x$ direction we have
\be
g^\mathrm{(sh)}_x=\alpha\,G\,(\rho_\mathrm{sh}-\rho_\mathrm{b})\,R\, \mathcal{I}_\mathrm{sh}\;,
\ee
with $\mathcal{I}_\mathrm{sh}$ different from $\mathcal{I}$ only by the limits of integration in $\xi$ and equal to
\be
\mathcal{I}_\mathrm{sh}=\int\limits^{2\pi}_0 \de \varphi\int\limits^{1+\frac{\Delta R}{R}}_1 \de \xi \;
\frac{\xi-\xi^2\cos\varphi}{\xi^2+1-2\xi\cos\varphi}=0\;.
\ee

\noi This result means that the spurious acceleration of the LG due to the part of the shell of compensation hidden behind the ZoA will also vanish if the density distribution $\rho_\mathrm{sh}$ is not constant, but depends on the distance from the centre of the LV (as we can divide the shell into infinitesimally thin layers of constant density each). The bottom-line is that a compensating shell with radial density distribution will \textit{not} affect the spurious acceleration of the Local Group provided that the latter is located at the inner edge of the shell, as observational constraints suggest.\footnote{In fact, the spurious acceleration from the hidden part of the shell would vanish even if the LG was \textit{inside} the LV, as long as the centre of the latter was coplanar with the Galactic equator.} On the other hand, if the LG was placed in the interior of the compensating layer, or at its outer edge, the discussed spurious acceleration from the LV would be \textit{diminished}, by up to $2\slash 3$ in the limit of an infinitesimally thin shell.

Having calculated the amplitude of the spurious acceleration in the model, we can apply our results to observational data. As an example, we take the analysis of \citet{Mall} of the clustering dipole of the 2-Micron All-Sky Survey (2MASS, \citealt{Skr}), since this is the densest and the deepest all-sky survey to date. \citet{Mall} used the measurements in the near-infrared $K_\mathrm{s}$ band and defined the Zone of Avoidance as the region with $|b|<7\degr$ for $l>230\degr$ or $l<130\degr$ and $|b|<12\degr$ for $l>330\degr$ or $l<30\degr$. Let us take a `mean value':
\be
\alpha=2\,\Delta b=\frac{1}{2}\,(2\times7\degr+2\times12\degr)=19\degr\simeq0.332\,\mathrm{rad}.
\ee
We now use Eq. (\ref{eq:v_spurious}) an for consistency with \citet{Tully.etal}, we apply \mbox{$H_0=74\kmsMpc$}, $\Omega_\mrm=0.24$ and $R=16\,\mathrm{Mpc}$, to obtain an approximate value of the additional, spurious velocity of the Local Group, measured if the Local Void is not properly accounted for:
\be
v_\mathrm{spur} \simeq 45\kms\;.  
\ee
When compared to the velocity of the LG relative to the CMB reference frame, $v_\mathrm{CMB}=622\pm35\;\kms$, one can see that this effect is of the same order as the error in the measurement of $v_\mathrm{CMB}$. Note also that due to the proportionality of $v_\mathrm{spur}$ to the radius of the LV (Eq. \ref{eq:v_spurious}), increasing its diameter to $45\,\mathrm{Mpc}$, while preserving sphericity, would cause the spurious velocity to raise significantly to $\sim60\kms$. However, owing to geometry of the problem, this would not largely affect general conclusions of our analysis (cf. subsequent Sections). Moreover, observational constraints point rather to some degree of elongation of the LV than to a larger size of the whole structure. We will now shortly address this issue.

The calculations so far assumed a simplistic model of a spherical void. But we can see for example on fig. 10 of \citet{Tully.etal} that the Local Void should be possibly modelled by a more sophisticated structure, like an ellipsoid. Current observational data suggest that the LV is elongated in the supergalactic $\mathrm{SGY}=0$ plane, roughly coincident with the Galactic plane, hence our $x$ axis. One should bear in mind however that this effect could be a manifestation of the ZoA itself: lack of observed galaxies in this region of the sky may be simply due to obscuration. Nevertheless, in case the elongation is real, for completeness of our analysis let us examine this possibility.

For that purpose we assume that the section of the LV in our $xy$ plane is an ellipse with semiaxes $b\equiv R$ and $a\equiv\kappa\,R $, where elongation \mbox{$\kappa>1$}. The major axis of the ellipse is placed along the $x$ axis of the coordinate system. As in the spherical case, owing to symmetries of the problem, the only non-vanishing component of the spurious acceleration `acting' on the LG is the $x$ one. It is easy to check that the relevant formula for $g_x$ is now
\be
g_x=-\alpha\,G\,\rho_\mathrm{b}\, \kappa \, R \, \mathcal{I}_e\;,
\ee
with the integral $\mathcal{I}_e$ given by
\be
\mathcal{I}_e=\int\limits^{2\pi}_0 \de \varphi\int\limits^{\Xi(\varphi)}_0 \de \xi \;
\frac{\xi-\xi^2\cos\varphi}{\xi^2+1-2\xi\cos\varphi}\;.
\ee
Here, $\Xi(\varphi)\equiv1\slash\left(\kappa\sqrt{1-e^2 \cos^2\varphi}\right)$ is a properly normalised equation of the ellipse in polar coordinates with eccentricity \mbox{$e\equiv\sqrt{1-\kappa^{-2}}$}.

The spurious velocity of the LG due to such elongation of the LV will be larger in comparison to the spherical case by a factor
\be
\tau\equiv\frac{v_\mathrm{spur} ^{(\mathrm{ell})}}{v_\mathrm{spur} ^{(\mathrm{sph})}}=\kappa\,\frac{\mathcal{I}_e}{\mathcal{I}}=\kappa\,\frac{\mathcal{I}_e}{\pi}\;.
\ee
Obviously, for $\kappa=1$, we have $\tau=1$. What is important here is that a linear increase in $\kappa$ results only in a slower than linear raise of the $\tau$ factor: for example $\kappa=2$ gives $\tau=4\slash 3$ and if $\kappa=3$, then $\tau=3\slash 2$. This means that for the observationally allowed elongation of $\kappa\simeq2$ and minor semiaxis $b\simeq15\Mpc$ \citep{Tully.etal}, the spurious velocity of the LV would raise by $1/3$ to \mbox{$\sim60\kms$}, which is the same as in the previously discussed case of the enlargement of a spherical and empty void. Note however that observations clearly show that the Local Void is not completely empty (cf. fig. 10 of \citealt{Tully.etal}) and this high value of the spurious velocity will be an upper limit for our considerations.

\section{Shift of the clustering dipole}
\label{Sec:Shift}
Knowing the amplitude of the spurious velocity induced by randomly-filled intersection of the Local Void and the Zone of Avoidance, we would like to check the shift of the \textit{direction} of the measured clustering dipole when the effect of the LV is accounted for. From Eq. (\ref{eq:g_spurious}), the true velocity of the LG (proportional to its acceleration in linear theory) is related to the calculated (`measured') one and the spurious component via
\be
\bmath{v}_\mathrm{T}=\bmath{v}_\mathrm{cal}- \bmath{v}_\mathrm{spur}\;.
\label{eq:vel_rel}
\ee

The vector $\bmath{v}_\mathrm{spur}$ is directed to the centre of the Local Void;\footnote{Note that although in our coordinate system the only non-vanishing component of the spurious velocity is the $x$ one, when projected on Galactic coordinates it has $x_\mathrm{Gal}$ and $y_\mathrm{Gal}$ components of comparable value, equal respectively to $0.87\,v_\mathrm{spur}$ and $0.5\,v_\mathrm{spur}$.} as $v_\mathrm{T}$ we adopt $v_\mathrm{CMB}=622\; \kms$. For the measured clustering dipole we choose the dipole of the 2MASS survey. Therefore, as the direction of the acceleration we use the values given in \citet{Mall} for random filling of the ZoA: $l_\mathrm{cal}=266\degr$, $b_\mathrm{cal}=47\degr$. Using these values altogether, after some calculations we find that the direction of the dipole is shifted down by $5\degr$ in $l$ and $2\degr$ in $b$. Thus, the `true' direction of the 2MASS dipole would be 
\be 
l_\mathrm{T} \simeq 261\degr\,, \qquad b_\mathrm{T} \simeq 45\degr\,.  
\ee
The shift is small, but as a systematic effect, it should be in principle accounted for in the measurement of the clustering dipole. However, random filling is not the only possible, nor the most optimal, way to treat the ZoA. A better method is to clone the sky below and above the ZoA, which has the advantage of tracing approximately structures through the ZoA. For the latter method, \citet{Mall} obtained $l_\mathrm{cal}=263\degr$, $b_\mathrm{cal}=40\degr$, so differences between the two methods give $\Delta l_\mathrm{cal}=3\degr$, $\Delta b_\mathrm{cal}=7\degr$. Therefore, the difference in the direction of the 2MASS dipole resulting from distinct methods of treating the ZoA is comparable to, or even greater than, the effect of the LV. These conclusions do not change significantly even if we include the elongation of the LV: for $v_\mathrm{spur}=60\kms$, we have a shift by $\Delta l=6\degr$ and $\Delta b=2\degr$ towards $l_\mathrm{T}=260\degr$, $b_\mathrm{T}=45\degr$.

The shift of the direction of the 2MASS clustering dipole changes the misalignment angle with respect to the CMB dipole. However, the amplitude of the shift is comparable to the uncertainty of the CMB dipole direction ($3\degr$ and $5\degr$ respectively for $l$ and $b$). Moreover, the calculated change of the misalignment angle turns out to be smaller than $1\degr$ even for high (but reasonable) values of $v_\mathrm{spur}$. We thus conclude that masking the LV has negligible effect on the misalignment angle between the 2MASS and CMB dipoles.

\section{Correcting the measured value of \hbox{$\bmath{\Omega_{\mathrm{\lowercase{m}}}}$}}
\label{Sec:Omega}

Application of Equation~(\ref{eq:dipole}) serves as a method to measure the cosmological parameter $\Omega_\mrm$ by comparing the velocity of the LG (equal to $\bfv_{\mathrm{CMB}}$) to its gravitational acceleration inferred from a galaxy survey. From Eq.~(\ref{eq:dipole}) it follows that $v_{\mathrm{CMB}} = \Omega_{\mrm}^{0.55}\, g_{\mathrm{LG}}$, where $g_{\mathrm{LG}}$ is the {\em scaled\/} gravitational acceleration of the LG (i.e. in units of velocity). The acceleration measured without the LV accounted for results in the `calculated' value of $\Omega_\mrm$, such that $\Omega_{\mathrm{cal}}^{0.55} = v_{\mathrm{CMB}}/g_{\mathrm{cal}}$. If the LV is taken into account, we find the `true' value of $\Omega_\mrm$, i.e. $\Omega_{\mathrm{T}}^{0.55} = v_{\mathrm{CMB}}/g_{\mathrm{T}}$. When we divide $\Omega_{\mathrm{cal}}^{0.55}$ by $\Omega_{\mathrm{T}}^{0.55}$, all the scaling factors relating the velocity to acceleration in linear theory cancel out. Therefore,

\be \label{eq:ratio}
\frac{\Omega_{\mathrm{cal}}^{0.55}}{\Omega_{\mathrm{T}}^{0.55}} =
\frac{v_{\mathrm{T}}}{v_{\mathrm{cal}}} \,.
\ee
The velocity $\bfv_{\mathrm{T}}$ is related to $\bfv_{\mathrm{cal}}$ by Equation~(\ref{eq:vel_rel}), where $\bfv_{\mathrm{spur}}$ is a small correction. Therefore, we can expect that the relative change in the value of $\Omega_{\mrm}$ will be small. We thus write $\Omega_{\mathrm{cal}} = \Omega_{\mathrm{T}} + \Delta\Omega$ and expand the expression $\Omega_{\mathrm{cal}}^{0.55} / \Omega_{\mathrm{T}}^{0.55}$ to first order in $\Delta\Omega/\Omega$. The result is \mbox{$\Omega_{\mathrm{cal}}^{0.55} / \Omega_{\mathrm{T}}^{0.55} = 1 + 0.55 \Delta\Omega/\Omega$}. The right-hand-side of Eq.~(\ref{eq:ratio}), calculated using Formula~(\ref{eq:vel_rel}) with $v_\mathrm{spur}=45\kms$, is $1.029$. Hence finally 
\be
\frac{\Delta\Omega_{\mrm}}{\Omega_{\mrm}} = 0.053 \,.  
\ee 
In other words, not accounting for the existence of the LV in measurements of the clustering dipole biases the estimated value of $\Omega_{\mrm}$ by about 5 per cent for the radius of the spherical LV equal to $16\Mpc$. If we allow for non-sphericity, this bias rises to some 7 per cent. This means that the influence of the LV on the determination of the cosmic density parameter from the comparison of the velocity and acceleration is small. Indeed, typically the uncertainty of the degenerate combination  $\beta\equiv f( \Omega_{\mrm})\slash b$, where $b$ is the \textit{linear biasing parameter}, amounts to at least $10\div20$ per cent (for recent determinations see e.g. \citealt{PH05} or \citealt{Erdogdu}). Owing to our ignorance of the exact value of $b$, we can conclude that the total relative error in $\Omega_{\mrm}$ is likely to be higher than in $\beta$ and the inclusion of the effect of the Local Void would not contribute largely to the total error budget, although it possibly should be included as a systematic effect.

The above analysis was performed within the linear theory, in which the estimated peculiar velocity, compared to the observed one, is inferred directly from the scaled peculiar acceleration (the clustering dipole). However, a completely empty void is a non-linear structure \citep{BiCh} and the actual spurious velocity of the LG, generated by the LV, will be greater than the corresponding scaled spurious acceleration. It is known that for a void with $\delta=-1$, the relation is $v_\mathrm{spur}\simeq1.5\,g_\mathrm{spur}$ (\citealt{B99}; \citealt{BiCh}). Application of this result would increase the systematic effects considered in this paper by roughly 50 per cent and enhance their significance. Nevertheless, in order to make this approach self-consistent, one would have to take into account non-linear effects from all other sources, especially those nearby, both over- and underdense. This would be a very difficult task, if not impossible, and is beyond the scope of this work, which deals with a simple model of the Local Void surrounded only by a shell of compensation.

\section{Summary and conclusions}
\label{Sec:Summ}
An important method to constrain the matter density parameter $\Omega_\mrm$ is to compare the peculiar velocity of the Local Group (known from the dipole component of the CMB temperature distribution) with its gravitational acceleration, inferred from all-sky surveys of galaxies (the \textit{clustering dipole}).

A serious problem plaguing such comparisons is that every survey called `all-sky' misses a significant amount of galaxies due to obscuration by dust, gas and stars in the disk of the Milky Way (the Zone of Avoidance, ZoA). To overcome this problem, in order to calculate the clustering dipole of the given survey, the ZoA is filled with mock galaxies. Their properties are chosen in a way to reflect the true, although unknown, galaxy distribution in the obscured part of the sky, basing on the one known from the rest of the celestial sphere.

A part of the ZoA intersects with a nearby void region, the Local Void (LV). When the existence of such a structure is not accounted for in the calculation of the acceleration of the LG, a spurious term is generated. In this paper we have calculated both the amplitude and the direction of this spurious acceleration. For simplicity we have assumed that the LV is spherical and for its size we have adopted the value estimated by \citet{Tully.etal}. We have also made the assumption that the LV is completely empty. Even then the amplitude of the spurious component amounts only to $45\kms$ in units of velocity. Including the observed elongation of the LV increases this value by $1/3$. On the other hand, possible presence of massive structures inside the LV, hidden behind the ZoA, could only lower this value.

This artificial acceleration changes also the direction of the calculated clustering dipole. We have shown that this change is comparable to the uncertainty in the direction of the peculiar velocity of the LG, determined from the dipole component of the CMB temperature distribution, reduced to the barycentre of the LG. Moreover, it points almost perpendicularly to the misalignment vector (i.e. the difference between the vectors of the velocity and acceleration of the LG). This results in a negligible shift of the misalignment angle, by less than one degree.

The final effect that we considered is the error in the inferred value of the non-relativistic matter density $\Omega_\mrm$ resulting from the negligence of the LV. We have estimated the relative error in this parameter as approximately $5\div7$ per cent. Therefore, up to this accuracy the influence of the Local Void on the determination of $\Omega_\mrm$ from velocity--density comparisons can be neglected. On the other hand, this additional biasing should be taken into account in the total error budget of the density parameter determined by such a method.

We would like to reiterate that our results do not negate the dynamical influence of the Local Void on the Local Group; on the contrary, \citet{Tully.etal} have shown that this influence is significant. It is only the effect of masking the intersection of the LV and the ZoA that seems to be of little importance for the purpose of calculation of the clustering dipole within the linear theory. This partially supports the claims that the Zone of Avoidance is not a crucial issue in determinations of the peculiar acceleration of the LG from all-sky surveys, especially such as 2MASS, where Galactic extinction is much weaker than in optical wavelengths.

\section*{Acknowledgements}
The authors would like to thank R. Brent Tully for the idea of this study and for useful comments on an earlier version of this manuscript. We also appreciate valuable suggestions from the re\-fe\-ree, Pirin Erdo\u{g}du.\\
\noi This work was partially supported by the Polish Ministry of Science and Higher Education under grant N N203 0253 33, allocated for the period 2007--2010.

\section*{} 

\end{document}